\documentclass[aps,prl,twocolumn,showpacs,groupedaddress]{revtex4}

\usepackage{epsfig}

\newcommand{\beq}{\begin{equation}}
\newcommand{\bea}{\begin{eqnarray}}
\newcommand{\eeq}{\end{equation}}
\newcommand{\eea}{\end{eqnarray}}

\begin{document}

\title{Critical Phenomena in Continuous Dimension}

\author{H.\ Ballhausen}
\author{J.\ Berges}
\author{C.\ Wetterich}
\affiliation{Institute for Theoretical Physics, Heidelberg University\\
Philosophenweg 16, 69120 Heidelberg, Germany\\}

\begin{abstract}
\noindent
We present a calculation of critical phenomena directly in continuous
dimension $d$ employing an exact renormalization group equation for the 
effective average action. For an Ising--type scalar field theory we 
calculate the critical exponents $\nu(d)$ and $\eta(d)$ 
both from a lowest--order and a complete first--order  
derivative expansion of the effective average action.
In particular, this can be used to study critical behavior as a function 
of dimensionality at fixed temperature.
\end{abstract}

\maketitle

\section{Introduction}

\noindent
One of the important quantitative successes of statistical physics is 
the precise understanding of critical phenomena based on the 
renormalization group~\cite{RG}. Universal properties near second--order 
phase transitions are found to be insensitive to the details of the 
underlying theory and only to depend on symmetry, field 
content and the dimensionality of space $d$. The crucial dependence of 
critical behavior on $d$ can be readily observed from the fact that no 
phase transition at nonzero temperature or, equivalently, no 
corresponding renormalization group fixed point 
exists for $d=1$. For higher dimensions phase transitions can occur, 
depending on the model. For instance, simple Ising--type one--component 
models can exhibit an infinite number of multi--critical points in 
$d=2$~\cite{Morris:1994jc}, 
a critical ``Wilson--Fisher'' or a tri--critical fixed point in 
$d=3$, and a trivial ``Gaussian'' fixed point for $d=4$. 

Continuity in the dimensionality of space is a crucial ingredient for some 
of the standard methods for the description of critical phenomena. 
In particular the extrapolation from $d=4$ to lower dimension 
($4-\epsilon$--expansion~\cite{RG}) is one of the most frequently
applied methods for studying the non-perturbative dynamics 
near a non-trivial fixed point of the renormalization group equation. 
In this language
the strong infrared divergences encountered by perturbation theory 
at fixed dimension are generated by an expansion 
around the ``wrong'' (Gaussian) fixed point. The  
$\epsilon$-expansion is used to interchange limits which
allows one to follow ``perturbatively'' the relevant 
fixed point. The involved extrapolation in the dimension $d$
is a crucial ingredient for this procedure. In particular,
the typical expansion parameter $\epsilon =1$ is not small
and improvement procedures such as Borel transformation and 
conformal mapping are required to obtain accurate 
results~\cite{ZinnJustin}. A direct calculation 
in continuous dimensions would provide important insight 
concerning the validity of such an approach. 

In this work we calculate critical phenomena for continuous 
dimension $d$ within a one-component scalar field theory with quartic 
self-interaction. Computations in arbitrary dimension can be 
performed~\cite{Aoki:1996fn,Tis} using the effective average action $\Gamma_k$ 
\cite{Wetterich:yh,Berges:2000ew}, which is infrared regulated by a 
momentum cutoff. The change of $\Gamma_k$ with the scale $k$ 
is described in terms of 
an exact flow equation~\cite{Wetterich:yh}:
\beq\label{EqexactRG}
\partial_k \Gamma_k[\phi] = \frac{1}{2} \textrm{Tr} 
\Big[ \left(\Gamma^{(2)}_k[\phi] + R_k\right)^{-1} \partial_k R_k \Big] \, ,
\eeq
where $\Gamma^{(2)}_k[\phi]$ is the exact inverse propagator 
and $R_k$ denotes the cutoff function. 
We investigate the theory 
in complete first--order in a derivative expansion, where
\beq\label{truncation}
\Gamma_k = \int {\rm d}^dx \Big[ U_k(\rho) 
+ \frac{1}{2} Z_k(\rho) \partial_\mu\phi \partial^\mu\phi \Big]
\eeq
with $\rho = \phi^2/2$. Critical phenomena 
can be described from the flow of the most general effective potential 
$U_k(\rho)$ consistent with the symmetries, as well as the wave 
function renormalization $Z_k(\rho)$~\cite{Berges:2000ew}.  
This approach allows one to perform direct calculations for a given continuous
dimension parameter $d$. In particular, this can be
used to study critical behavior as a function of 
$d$ at fixed temperature.

\section{Evolution equations}

For the truncation (\ref{truncation}) the exact renormalization group 
equation (\ref{EqexactRG}) is reduced to partial differential equations 
for $U_k(\rho)$ and $Z_k(\rho)$. These equations can be found in 
Refs.~\cite{Berges:2000ew,VonGersdorff:2000kp}. For the cutoff function 
we employ~\cite{Litim:2001up}  
\beq
R_k= Z_k\, (k^2 - p^2)\, \Theta(k^2 - p^2)
\label{thetacut}
\eeq
with $Z_k = Z_k(\rho=\rho_0)$ evaluated at vanishing external momentum. 
The flow of the function $Z_k$ is determined by the anomalous dimension 
$\eta = - \partial_t \ln Z_k$. Our quantitative results are based on a 
complete numerical solution of the equations for the first--order derivative 
expansion. For comparison, we solve in addition a simpler approximation with a 
field independent wave function renormalization $Z_k$. For our analytical 
discussion we concentrate on the latter, for which the flow of the rescaled 
effective potential $u(\tilde{\rho}) = k^{-d} U_k$ as a function of 
the renormalized and dimensionless field variable 
$\tilde{\rho}= Z_k k^{2-d} \rho$ is described by
\bea
\partial_t u &=& - d u + (d-2+\eta) \tilde{\rho} u^\prime 
\nonumber\\
&& + \frac{4 v_d}{d} \left(1 - \frac{\eta}{d+2} \right)
\frac{1}{1+u^\prime+2\tilde{\rho} u^{\prime\prime}} \,\, ,
\label{flowu}
\\[0.2cm]
\label{floweta}
\eta &=& \frac{8 v_d}{d} 
\frac{\kappa (3\lambda+2\kappa u_3)^2}{(1+2\kappa \lambda)^4} \,\, ,
\eea
with $v_d^{-1} = 2^{d+1} \pi^{d/2} \Gamma(d/2)$ and 
$u^\prime=\partial u/\partial \tilde{\rho}$. Here
$t=\ln(k/\Lambda)$, where $\Lambda$ denotes a high momentum scale. 
Most of the relevant qualitative properties can be obtained 
from the $k$-dependence of the location $\kappa$ of the minimum of 
$u(\tilde{\rho})$. By taking the total $t$-derivative of 
$u'(\tilde{\rho}=\kappa)=0$ 
the scale-dependence of the minimum for $\kappa > 0$ is 
found to be:
\bea
\label{betakappa}
\partial_t \kappa &=& 
\beta_\kappa = - ( d - 2 + \eta ) \kappa + \frac{12 v_d}{d} N_r \, ,
\\[0.2cm]
\label{effN}
N_r &=& \left(1 - \frac{\eta}{d+2}\right)
\frac{1+2\kappa u_3/(3\lambda)}{(1 + 2 \kappa \lambda)^2} \, .
\eea
Here the couplings are related to $\tilde{\rho}$-derivatives of 
the rescaled potential evaluated at the minimum, 
i.e.\ $\lambda = u''(\kappa)$ and $u_3 = u'''(\kappa)$. 
Note that $\beta_\kappa$ is parametrized in terms of the 
couplings $\lambda$ and $u_3$. Implicitly, however, 
$\beta_\kappa$ depends on an infinite number of couplings 
since the correponding $\beta$--function for $\lambda$ 
depends on the higher couplings $u_3$ and 
$u_4=u^{(4)}(\kappa)$, the one for $u_3$ depends on $u_4$ and 
$u_5=u^{(5)}(\kappa)$ etc. 

A second--order phase transition is characterized by a fixed point 
or scaling solution for which \mbox{$\partial_t u^\prime=0$} and, 
in particular, $\beta_\kappa=0$. All renormalized dimensionless couplings 
become independent of the scale $k$ and take on their 
''fixed point values'' $\kappa_*$, $\lambda_*$, $u_{3*}$ etc.
Typically, $\kappa$ corresponds to the relevant parameter 
such that critical behavior can be observed only for a 
fine--tuned initial value 
$\kappa(\Lambda) \simeq \kappa_c$ with 
$(\kappa(\Lambda) - \kappa_c) \sim (T_c-T)$. 
If $\kappa$ is the only relevant parameter 
the flow of all other couplings $\lambda$, $u_3$ and 
higher derivatives of the potential follow fixed 
''critical trajectories'' $\lambda(\kappa,d)$, $u_3(\kappa,d)$ 
--- so--called partial fixed points. 
(More precisely, this holds if $T$ is in the vicinity of the 
critical temperature $T_c$ and $k$ is sufficiently below $\Lambda$.)
Inserting the critical trajectories, $\beta_\kappa$ becomes only 
a function of $\kappa$ and $d$: 
\beq 
\beta_\kappa(\kappa,d) \equiv 
\beta_\kappa(\kappa,\lambda(\kappa,d), \ldots, d) \, . 
\eeq
For arbitrary $d$ the critical behavior is 
then directly linked to the properties of
$\beta_\kappa(\kappa,d)$ --- in particular its zeros. The 
precise form of the critical trajectories
$\lambda(\kappa,d)$ etc. is often not known analytically 
but it can be determined numerically. However, we
will see that many
crucial properties of $\beta_\kappa(\kappa,d)$
do actually not involve a very precise analytical understanding 
of the critical trajectories. We emphasize that the use of
$\lambda(\kappa,d)$ etc.~in the flow equations is valid only in the 
vicinity of the critical trajectory. The analytical
discussion in the next section is therefore limited to the behavior 
close to criticality. The full numerical treatment below is
not restricted to this.  

\section{Asymptotic behavior for large $\kappa$}

As a general feature, $\beta_\kappa(\kappa,\lambda,...,d)$ 
is analytic in all 
arguments except for $\lambda=0$. This does, however, not imply 
automatically that 
$\beta_\kappa(\kappa,d)$ is analytic since the solutions for the
critical trajectories of
$\lambda(\kappa,d)$ and the higher couplings can be non--analytic. 
For fixed $\kappa$, 
$\lambda$, etc.\ the function $\beta_\kappa$
decreases with increasing $d$. According to (\ref{betakappa}), 
(\ref{effN}) and (\ref{floweta}) one has
\beq
\beta_\kappa = \frac{12 v_d}{d} \quad \textrm{for} ~ \kappa=0 \, .
\eeq
To discuss $\beta_\kappa$ for $\kappa \to \infty$
we note that the leading behavior is given by
\begin{equation} 
\beta_\kappa = \Bigg\{ \matrix{ (2-d)\kappa 
&& \textrm{for} ~ \kappa \to \infty, ~ d > 2 \cr \vspace*{0.1cm}    
c_\infty^{(d)}  &&  \textrm{for} ~ \kappa \to \infty, ~ d \le 2} \,\,\, .
\label{beta}
\end{equation} 
As is shown below, here $c_\infty^{(d)}$ is a constant which depends on $d$.
In particular, in two dimensions one finds 
\mbox{$c_\infty^{(2)}=-1/(4\pi)$}~\cite{VonGersdorff:2000kp}.
For positive $c_\infty^{(d)}$ the asymptotic behavior
can be related to a so--called ``zero--temperature phase transition''.
In this case critical behavior can be associated with a fixed point for
$\partial_t \kappa^{-1} = - \kappa^{-2} \beta_\kappa$ 
at $1/\kappa=0$. 

To establish the leading behavior (\ref{beta}) 
for $d<2$ we first note that the region of large $\kappa$ admits
slowly evolving solutions if $\eta$ is near $2-d$. 
From Eqs.~(\ref{betakappa}) and (\ref{floweta}) one observes
\beq\label{F1}
\kappa_*^3\lambda_*^2=\frac{2v_d}{d(d+2)^2}\, \eta
\Big(\frac{d+2-\eta}{d-2+\eta}\Big)^2  \, ,
\eeq
which shows that the presence of fixed points in the region of 
large $\kappa^3\lambda^2$ requires values of
$\eta$ close to $2-d$.
We parametrize $\eta=2-d+\zeta/\kappa$, where $\zeta$ is a
$\kappa$-independent constant. Introducing the notation
$u(\tilde{\rho})=\tilde{\rho}^{-1/2} h(\tilde{\rho})$
one finds from Eq.~(\ref{flowu}):
\bea
 \partial_t h \mid_{\tilde{\rho}} &=& -dh - \frac{\zeta}{2\kappa} h
+ \frac{\zeta}{\kappa} \tilde{\rho} h^\prime \\
& & \nonumber + \frac{4v_d}{d+2}\Big(1-\frac{\zeta}{2\kappa d}\Big)
\frac{1}{h^{\prime\prime}-\frac{h^\prime}{2\tilde{\rho}}
+ \frac{h}{2\tilde{\rho}^2} +\frac{1}{2\sqrt{\tilde{\rho}}}}  \, .
\label{G1}
\eea
Since we are interested in values for 
$\tilde{\rho}$ near $\kappa$ we consider the variable 
\mbox{$x=\tilde{\rho}-\kappa$} and neglect terms suppressed 
by powers of $x/\kappa$ as well as $\zeta/\kappa$, i.e.
\bea
 \partial_t h \mid_x &=& \partial_t h \mid_{\tilde{\rho}} 
+ h^\prime \partial_t \kappa \nonumber\\	
&=& -dh 
+ \frac{12v_d}{d}N_r h^\prime 
+ \frac{4 v_d}{(d+2)h^{\prime\prime}}
- \zeta h^\prime \, .
\label{G2}
\eea
We observe that $\kappa$ does not appear anymore in (\ref{G2}) 
and consequently quantities like 
$h_2 \equiv h^{\prime\prime}(x=0)$, 
$h_3 \equiv h^{\prime\prime\prime}(x=0)$
are not expected to depend on $\kappa$. 
In leading order for $\kappa \to \infty$ we therefore have 
$\lambda\sim\kappa^{-1/2}$, $u_3\sim\kappa^{-1/2}$ 
and $\lambda\kappa\sim\kappa^{1/2}\gg1$
as well as $\kappa u_3\sim\kappa^{1/2}\gg\lambda$. 
In particular, $N_r$ is also independent of $\kappa$ and given by
\beq
N_r=\frac{d}{3(d+2)}\,\frac{h_3}
{h_2^3}  \, .
\label{G3}
\eeq
This shows that the leading behavior of $\beta_{\kappa}$ for
$\kappa \to \infty$ and $d<2$ is described by     
\beq
\beta_\kappa=-\zeta+\frac{12v_d}{d}N_r=c_\infty^{(d)}  \, .
\label{G7}
\eeq

Further information can be obtained from 
the scaling solution $\partial_t h|_x =0$.
(Note that the scaling solution for 
$h$ does not correspond to an exact fixed point since 
$\kappa$ evolves according to (\ref{G7}).)
At the potential minimum we have $h^\prime(x=0)=0$ 
up to corrections $\sim 1/\kappa$. From (\ref{G2}) one then finds
\beq 
h_2=\frac{4v_d}{d(d+2)h_0} \, .
\eeq
The integration constant $h_0=h(x=0)$ has to be chosen 
such that the lowest order relation for the anomalous
dimension, $\eta_0 = 2 - d$, holds. According to
Eq.~(\ref{floweta}) at this order
\beq
\eta_0 = \frac{18 v_d}{d^3} (d+2)^2 h_2^2 N_r^2 \stackrel{!}{=}2-d \, ,
\label{G5}
\eeq
which fixes $N_r$ as a function of $h_2$:
\beq
N_r=\pm\sqrt{\frac{(2-d)d}{2v_d}}\,\frac{d}{3(d+2) h_2}
\label{G6}   \, 
\eeq
and $h_3$ by (\ref{G3}).
Finally, $h_2$ is constrained by the requirement 
that the scaling solution should extend to the whole 
range $-\kappa<x<\kappa$. For $d \to 2$ one concludes 
from (\ref{G6}) that $N_r \to 0$. As a consequence,
$h$ is determined by the motion of a particle
in a logarithmic potential, i.e.~for $d \to 2$ 
the lowest order scaling solution for (\ref{G2}) reads
\beq
(h')^2 = \frac{8 v_d}{d (d+2)} \ln \left(\frac{h}{h_0} 
\right) \, .
\eeq

It is striking that the simple equations (\ref{flowu}),
(\ref{floweta}) and  
(\ref{betakappa}) reproduce correctly the low--temperature 
behavior of the exact Ising model for $d=1$, which is
characterized by $\eta=1$
and essential scaling: For $T\to 0$ from the leading behavior 
(\ref{G7}) one has
\beq
\frac{\partial \kappa}{\partial k} = \frac{c_\infty^{(1)}}{k} \, .
\eeq
Associating the scale $k_s$ where 
$\kappa(k_s)=0$ to the inverse correlation length $\xi^{-1}$ up to a factor 
one finds
\beq
\xi\Lambda\sim\exp\Big(
\frac{\kappa(\Lambda)}{c_\infty^{(1)}}\Big)\sim\exp(2J/T)
\label{G8}
\eeq
with $\kappa(\Lambda)=b/T$, $b=2c_\infty^{(1)}J$.

\section{Non-zero temperature phase transition}

From the asymptotic behavior discussed above one can draw a number
important conclusions about the presence of a
phase transition at non-zero temperature, 
characterized by a {\it finite} value $\kappa_{*}$. 
We note that a positive value of $\beta_\kappa$ for 
$\kappa=0$ and negative values for large $\kappa$
require the $\beta$--function to vanish at an intermediate 
$\kappa=\kappa_*$. Since \mbox{$c_\infty^{(2)}=-1/(4\pi)$}
Eq.~(\ref{beta}) immediately tells us that for all $d\ge 2$
a finite value $\kappa_{*}$ exists for which $\beta_\kappa=0$.
For instance, in $d=3$ the corresponding phase 
transition is described by the well-known Wilson--Fisher fixed 
point in the presence of only one relevant parameter~\cite{ZinnJustin}. 

From the exact solution of the one-dimensional Ising--model 
one has $c_\infty^{(1)}>0$ and $\beta_\kappa$ is above a positive 
constant for all $\kappa$. Therefore, no fixed point with a 
finite $\kappa_*$ occurs in $d=1$. If the $\beta$--function depends
continuously on $d$ we conclude that the zero-temperature 
transition and the absence of a fixed point with finite $\kappa_*$ 
extend to a finite range of $d > 1$. (It is crucial
in this respect that the limiting value $c_\infty^{(1)}$ is 
strictly positive.) As a consequence,
the lower critical dimension, below which phase transitions
at nonzero temperature are absent, has to be larger than one. 

The lower bound on the critical dimension, $d_{\rm low} >1$,
can be supplemented by an upper bound using similar continuity
arguments. For this we note that
the correlation length exponent $\nu$ 
at an infrared--unstable fixed point is given by the first derivative 
of $\beta_\kappa(\kappa,d)$ at $\kappa_* \neq 0$:
\beq
\nu^{-1} = - \frac{\partial \beta_\kappa(\kappa,d)}{\partial\kappa}
\Big|_{\kappa_*} 
\, .
\eeq
For finite $\nu$ this yields the usual power law for the
correlation length, 
$\xi \sim |T-T_c|^{-\nu}$~\cite{ZinnJustin}. 
For a vanishing first derivative the exponent $\nu$ diverges. In this
case the power law is replaced by essential scaling. 
A finite value for $\nu$ corresponds to nonzero negative 
$\partial\beta_\kappa/\partial\kappa_{\mid\kappa_*}<0$. Since the 
derivative of the $\beta$--function
(and therefore $\nu$) depends continuously on $d$, 
we conclude that for finite $\nu$ 
at a given $d$ also a finite neighborhood in $d$ 
exhibits a phase transition. At the lower critical dimension,
below which no phase transition with a finite $\kappa_*$ occurs, 
either $\kappa_*\to\infty$ or $\nu$ must diverge and $d_{\rm low}$ 
is characterized by essential scaling. 
The finiteness of $\nu$ for $d=2$ follows from 
universality and the exact value for
the Ising model, $\nu=1$~\cite{ZinnJustin}. Therefore $\nu$ must also 
remain finite in a finite range
$d<2$. This continuity argument therefore implies $d_{\rm low} < 2$. 

\begin{figure}[t]
\begin{center}
\vspace*{0.3cm}
\hspace*{-3.cm}
\epsfig{file=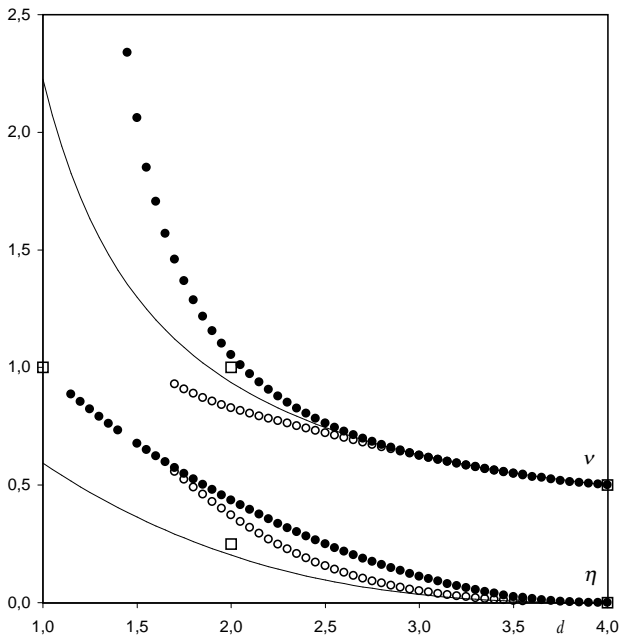,width=8.cm}
\end{center}
\vspace*{-1.7cm}
\caption{The critical correlation length exponent $\nu$ and 
anomalous dimension $\eta$ as a function of the dimension $d$ for
two different approximations: complete first-order derivative expansion
(open circles) and for a field-independent wave function renormalization
(full circles). We also show the results
of the $\epsilon$-expansion up to $\epsilon^5$ (solid line) [8],
which are plotted for the range $0 \le \epsilon \le 3$ for
illustration, along with
available exact results (open squares).}
\label{fig2}
\end{figure}
We emphasize that for a second--order phase transition,
apart from a vanishing $\beta_\kappa$--function, all couplings 
have to take on fixed--point values and the critical
trajectories $\lambda(\kappa,d)$ etc.\ have to be well defined.
In particular, the character of a fixed point is influenced
by the stability properties with respect to the other couplings. 
To establish this we turn to the numerical solution of the full 
first--order derivative expansion, as well as for the approximation
with a field-independent wave function renormalization
as discussed above. Since the approximations include a
most general potential term $u(\tilde{\rho})$ the description
takes into account an ``infinite number of couplings'' 
$u_n = u^{(n)}(\tilde{\rho})|_{\tilde{\rho}=\kappa}$. 

We numerically solve the evolution equations, which are 
given e.g.~in Eqs.~(3.8) and (3.10) of Ref.~\cite{Wetterich:2001kr},
for the one-component theory to first order in the derivative expansion.
The equations were discretized on a grid for the variable $\tilde{\rho}$
with varying grid-size (typically thirty points and much higher resolution
in lower dimensions).
The resulting ordinary differential equations
were solved using a fifth-order Runge-Kutta
algorithm and a fourth-order Cash-Karp
stepping routine. The relative local errors of the
integration routine were below $10^{-6}$.
Results for sample exponents in various dimensions
were cross-checked with different numerics (in particular,
the integration of the flow equation of the 
potential $u(\tilde{\rho})$ instead of 
its first derivative $u'(\tilde{\rho})$, as well as a fixed grid instead of
running discretization points $\sim \kappa$).
The quantitative differences in the exponents 
using the different numerics were \mbox{manifest} only 
in the third to fifth significant digit. 

The results for the critical exponent
$\nu(d)$ and the anomalous dimension $\eta(d)$ are plotted
in Fig.~\ref{fig2} for the two different employed approximations.
For comparison we also include the results obtained from the
expansion around $4 - \epsilon$ dimensions up 
to $\epsilon^5$~\cite{Kleinert:rg,ZinnJustin}
as well as available exact results. 
The results from the $\epsilon$-expansion have been plotted 
for values as large as $\epsilon = 3$ for illustrational 
purposes though it is typically employed only for
$\epsilon = 1$. In contrast, we emphasize that our results correspond 
to direct calculations for the corresponding dimensions. In 
particular, they do not involve any improvement procedure
such as Borel transformation and conformal mapping employed for
the expansion around $4 - \epsilon$~\cite{Kleinert:rg,ZinnJustin}.

To numerically obtain $\nu(d)$ we calculate a series
of near--critical trajectories with varying
$(\kappa(\Lambda) - \kappa_c) \sim (T_c-T)$. These calculations 
become time intensive for low dimensions, and grow
substantially for the more sophisticated approximation with a 
field-dependent $Z_k(\tilde{\rho})$. This is why we have computed
$\nu(d)$ only down to about $d=1.5$ in Fig.~\ref{fig2}.  
In contrast, the anomalous dimension $\eta$ can be directly
infered from the critical trajectory and is presented in Fig.~\ref{fig2}
for even lower dimensions. 
 
As $d$ is lowered from the upper critical dimension
$d=4$ both $\nu(d)$ and $\eta(d)$ grow. One observes a
very good agreement of the results from our first-order derivative 
expansion and the $\epsilon$-expansion results down to about 
$d=3$. At $d=2$ we can compare with the exact values $\nu = 1$ 
and $\eta = 0.25$. The renormalization group results for $\nu$ with the
$\Theta$-cutoff function (\ref{thetacut}) show a rather strong
dependence on the approximation around $d=2$. 
The strong increase of $\nu(d)$ as $d$ is further lowered sets in 
much earlier for the approximation with a field-independent 
wave-function renormalization. Interestingly,
for the anomalous dimension the dependence on the approximation is diminished
for lower dimensions. In particular, we observe that already the less
sophisticated approximation approaches the exact one-dimensional result. 
This may be related to fact that in one dimension the propagator 
shows a simple quadratic dependence on momenta for $\eta = 1$~\footnote{
The exact propagator $G$ of the one-dimensional Ising model in the 
zero temperature limit is given in Fourier space by 
$G(q) = f(q \xi )/q^{2-\eta} = f(q \xi )/q
\sim \xi^{-1}/(q^2 + \xi^{-2})$ 
with $f(q \xi) \sim (q\xi)^{-1}/(1+ (q \xi)^{-2})$.
}. 

\section{Critical behavior as a function of dimensionality}

We point out that critical
behavior can also be studied as a function of $d$ for {\it fixed} 
$\kappa(\Lambda)$ or, equivalently, temperature $T$. 
In the critical region we find that for fixed $T$ the 
correlation length varies with $d$ as
\begin{equation} 
\xi \sim |d_c-d|^{-\nu}  \, .
\end{equation} 
Here $d_c$ is the dimension for which $T$ becomes the critical
temperature and $\nu = \nu(d_c)$ is the {\it same} critical
exponent as computed above for the variation with $T$ for
fixed $d$.
This property can be understood as a simple consequence 
of continuity. Assume some fixed point
where $\beta_\kappa(\kappa_*(d_c),d_c)=0$ and 
$(\partial\beta_\kappa/\partial\kappa)
(\kappa_*(d_c),d_c)=-\nu^{-1}(d_c)$
with finite critical exponent $\nu(d_c)$. 
Starting with critical initial values 
$(\kappa_*(d_c),d_c)$ at some scale 
$\bar{\Lambda}$ the correlation length $\xi$ diverges. 
For initial values 
$(\kappa_*(d_c)-\Delta,d_c)$ we have the 
well--known critical behavior
($\Delta>0$ in the symmetric phase)
$\xi(\Delta,d_c) \sim \Delta^{-\nu(d_c)}$.
Continuity tells us that for $d=d_c-\Delta_d$ there is
a new fixed point with 
$\beta_\kappa(\kappa_*(d_c-\Delta),d_c-\Delta)=0$. 
Instead of varying
$\Delta$ at fixed $d_c$ one may keep a fixed 
$\kappa(\Lambda)=\kappa_*(d_c)$
and vary $\Delta_d$. With $\kappa_*(d_c)
=\kappa_*(d_c-\Delta_d)-\tilde{\Delta}$
this yields
$\xi(0,d_c-\Delta_d) 
\sim \tilde{\Delta}^{-\nu(d_c-\Delta_d)}$.
(Note that $\Delta_d>0$ for the symmetric phase.) For 
$\Delta_d\to 0$ we can relate 
$\tilde{\Delta}$ to $\Delta_d$
\begin{equation} 
\tilde{\Delta} 
= - \frac{\partial \kappa_*}{\partial d}
(d_c)\Delta_d = - \nu(d_c) 
\frac{\partial \beta_\kappa}{\partial d}
(\kappa_*(d_c),d_c)\Delta_d  \, .
\end{equation}
Generically one has $\partial \beta_\kappa / \partial d > 0$ 
and we conclude that the dependence
of $\xi$ on $d$ at the critical point $\kappa_*(d_c)$ 
is characterized by the standard
critical exponent $\nu=\nu(d_c)$,
which agrees with the results from the numerical solution 
to very good accuracy. This is investigated for the corresponding
$N$--component model in Ref.~\cite{Ballhausen}. We recall, however,
that this argument is only valid for finite $\kappa_*$ and away 
from essential scaling at $d_{\rm low}$.

\section{Conclusions}

We have presented a direct calculation of critical phenomena 
in continuous dimension emloying a complete first-order
derivative expansion. The results for the critical exponents
show good agreement with those found from the $\epsilon$-expansion
for not too small dimensions down to about $d=3$. The deviations
grow substantially for lower dimensions. To get an idea about the
reliability of our results we have re-calculated the exponents using
a less sophisticated approximation with a field-independent wave-function 
renormalization. It is striking
to observe that though the critical
behavior depends comparably strongly on the employed 
approximation scheme for intermediate values of $d$, this has 
only a small effect on some nontrivial properties at 
sufficiently low dimensions: even the less sophisticated approximation 
is sufficient to approach the exact result for the anomalous dimension 
$\eta =1$ in $d=1$. It has recently been demonstrated for the
three-dimensional one-component model that the next
order ($\partial^4$) of the derivative expansion leads to significantly 
improved results~\cite{Canet:2003qd} compared with lower orders calculations. 
It would be very interesting to investigate the apparent convergence 
of the derivative expansion for lower dimensions as well.
 
We have pointed out that with the employed methods one can use
the dimensionality $d$ as a ``relevant'' parameter to study
deviations from criticality. For fixed temperature $T$
the deviations from the critical dimension $d_c$ are described
by the same universal critical exponents which parametrize the 
deviations from $T_c$ for fixed $d$. So far we are not aware of
any clear experimental application of the calculated critical phenomena 
in fractional dimensions. However, an effectively similar behavior may 
be observed e.g.~for the magnetic properties of thin films, showing a 
dimensional crossover from three-- to two--dimensional scaling 
properties by varying the film thickness \cite{thinfilms}.

\bigskip
\noindent
We thank Matthieu Tissier for helpful discussions.

\end{document}